\documentclass[aps,prd,reprint,onecolumn,notitlepage,superscriptaddress]{revtex4-1}

\usepackage{lipsum}
\usepackage{amsmath}
\usepackage{breqn}
\usepackage{graphicx}
\usepackage{color}
\usepackage{tikz}
\usepackage{verbatim}
\usepackage{hyperref}
\usepackage{amssymb}
\usepackage{bm}
 \usepackage{booktabs} % For professional-looking tables
\usepackage{multirow} % To span rows
\usepackage{caption}
\usepackage{subcaption}
\usepackage{comment}
\newcommand{\rom}[1]{\textup{\uppercase\expandafter{\romannumeral#1}}}

\begin{document}

%\title{Exploring Coupled Quintessence scenario with Warm Dark Mater}
\title{Exploring Coupled Quintessence in light of CMB and DESI DR2 measurements}
\author{Atul Ashutosh Samanta\footnote{satulashutosh@gmail.com}
       }

\affiliation{Indian Institute of Science Education and Research$,$ Bhopal$,$ 462066$,$ India}

\author{Abhijith Ajith \footnote{abhijith.ajith.421997@gmail.com}
       }

\affiliation{Indian Institute of Science Education and Research$,$ Bhopal$,$ 462066$,$ India}

\author{Sukanta Panda \footnote{sukanta@iiserb.ac.in}
       }

\affiliation{Indian Institute of Science Education and Research$,$ Bhopal$,$ 462066$,$ India}

\begin{abstract}
We perform a detailed analysis of a theoretically motivated dark energy quintessence model which interacts with the dark matter sector of the universe. Utilizing the current observational datasets from the Cosmic Microwave Background, Baryon Acoustic Oscillations and Type Ia Supernovae, we constrain the parameters that characterize the strength of the time dependent interaction. We also look at the effect of a warm dark matter component in the context of coupled quintessence. Analysis using Deviance Information Criterion indicates strong preference for the quintessence model coupled with warm dark matter. However, Bayesian evidence analysis shows favor in the direction of $\Lambda$CDM model.

\end{abstract}
\maketitle

\section{Introduction}
\label{intro}
Over the last two decades, cosmology has undergone unprecedented changes thanks to the advancements in precision technologies, driven by a wealth of high-precision observational data. The observations from the Cosmic Microwave Background (CMB) \cite{Planck:2018vyg,Planck:2019nip,Planck:2018lbu, SPT-3G:2021eoc, ACT:2020frw, WMAP:2012nax, Planck:2018nkj,Lemos:2023rdh,Tristram:2007zz}, Baryon Acoustic Oscillations (BAO) \cite{BOSS:2016wmc,eBOSS:2020tmo,eBOSS:2020gbb,eBOSS:2020yzd,eBOSS:2020hur,eBOSS:2020lta,DESI:2025fxa,DESI:2024mwx}, Type Ia supernovae \cite{Pan-STARRS1:2017jku,Scolnic:2021amr,Rubin:2023ovl,Bailey:2022pax,Blondin:2024fpr,DES:2024jxu,NearbySupernovaFactory:2015pcf} and weak lensing \cite{DES:2016jjg,DES:2018ekb,LineaScienceServer:2021mgv,Hildebrandt:2016iqg,Hildebrandt:2018yau,HSC:2018mrq,DES:2022ccp,DES:2021wwk}, have significantly enhanced our understanding of the universe. The standard cosmological model points out that the majority of the universe's energy content consists of two mysterious components: Dark matter (DM), that drives the formation of large-scale structure, and Dark energy (DE), which is believed to be responsible for the observed accelerated expansion of the universe. The nature of these components has been a subject of constant debate, with Cold Dark matter (CDM) and Cosmological constant ($\Lambda$) being popular candidates in the scientific literature \cite{Blumenthal:1984bp,Peebles:1982ff,Efstathiou:1990xe,Peebles:1984ge}. The $\Lambda$CDM model fits well with a wide range of observations. However, with the improvement of measurement precision over the years, several tensions have appeared between the probes of the early and late universe \cite{Verde:2019ivm,Hildebrandt:2016iqg,DES:2017myr,Planck:2018vyg}, which question the underlying nature of the universe's components. The Hubble tension refers to the inconsistency between the measurements of the current expansion
rate of the Universe ($H_0$), measured directly  using the probes in the late time, and those measurements for which the values relied on the inference of the $\Lambda$CDM cosmological model calibrated on probes from early universe. To be precise, the $H_0$ tension refers to the 4 to 6$\sigma$ discrepancy
between the $H_0$ measurements from Planck ($H_0 = 67.36 \pm 0.54$ km/s/Mpc \cite{Planck:2018vyg}) and the late-time measurements from the SH0ES collaboration ($H_0 = 73.04 \pm 1.04$ km/s/Mpc \cite{Riess:2021jrx}). The $S_8$ tension refers to the discrepancy between the value  of the matter fluctuation amplitude parameter $S_8$ obtained from 
the CMB by Planck \cite{Planck:2018vyg}, $S_8 = 0.834 \pm 0.016$, and the directly measured values from KiDS-1000 \cite{Hildebrandt:2016iqg} and DES Y3 \cite{DES:2021wwk}, $S_8= 0.769\pm0.016$, indicating a 2.9$\sigma$ deviation. Such disagreements in the context of the $\Lambda$CDM have motivated the exploration of new candidate models. 

 Dark matter models include candidates such as WIMPs, axions, sterile neutrinos, non-cold dark matter and so on \cite{PhysRevD.89.041702,OHare:2024nmr,Boyarsky:2018tvu,2016PhRvD..94h3504C,Kumar:2025etf}. These candidates try to explain many astrophysical and cosmological observations such as the flat rotation curves of galaxies, the dynamics of galaxy clusters, the gravitational lensing of background objects, the formation of large-scale structure, and so on that cannot be fully accounted for by baryonic matter alone. Similarly,  many promising candidates for dark energy such as quintessence, K-essence, Chaplyn gas, Holographic dark energy and so on have been of popular interest in the scientific community. The strong preference for an evolving dark energy from the DESI results \cite{DESI:2024mwx,Cortes:2024lgw}, has prompted the surge in the exploration of a plethora of dark energy models \cite{PhysRevD.37.3406,Chiba:1997ej,PhysRevD.62.023511,PhysRevD.63.103510,Sotiriou:2008rp,Manoharan:2022qll,Rinaldi:2014yta,Hu:2007nk,Casalino:2018tcd,Mukhopadhyay:2019cai,Ruchika:2020avj,Odintsov:2020zct,Ong:2022wrs,Luciano:2023wtx,Tyagi:2025zov,Akrami:2025zlb,Santos:2025wiv,Colgain:2025nzf,Artymowski:2020zwy,Ben-Dayan:2023rgt,Ben-Dayan:2023htq}, and DE-DM interactions \cite{Amendola:1999er,An:2017crg,Yang:2017yme,DiValentino:2019ffd,Patil:2023rqy,vanderWesthuizen:2023hcl,Ashmita:2024ueh, Gomes:2023dat}. 

Quintessence, originally introduced as a dynamical scalar field model for dark energy \cite{1988ApJ...325L..17P}, has been widely studied not only in isolation but also in scenarios where it interacts directly with the dark matter sector. Countless observations support the theory of dark sector interaction, also constraining the transfer of energy from DM to DE or vice versa \cite{Barreiro:2010nb,Yang:2018pej,An:2017crg,Yang:2017yme,Yang:2014okp,Fay:2016yow,Piloyan:2014gta,Li:2013bya,Yang:2018euj,DiValentino:2019ffd,DiValentino:2020kpf,Gomez-Valent:2022bku}. Further, they indicate the impact of the coupling on the evolution of the universe \cite{Jackson:2009mz,Zimdahl:2001ar}. The effects of such interactions on the CMB and linear matter power spectrum \cite{Linton:2017ged,Olivares:2006jr}, the structure formation \cite{Valiviita:2008iv,Duniya:2015nva,Carbone:2013dna}, halo mass function \cite{Tarrant:2011qe} and on the behavior of the cosmological parameters \cite{He:2008tn,Binder:2006yc,Patil:2022ejk,Patil:2022uco} have been well investigated. Apart from these, coupled quintessence models have been studied in various other contexts to provide viable cosmological solutions for the same \cite{vandeBruck:2016jgg,Koivisto:2009ew,Koivisto:2009fb,Koivisto:2012xm,Barros:2015evi,Gomes:2015dhl}. They have also been studied extensively to alleviate the Hubble tension. However, theoretical studies on such models reveal the appearance of negative energy densities and big rip scenario with them, but can be avoided by applying suitable conditions on the interaction strength and direction of energy transfer \cite{vanderWesthuizen:2025rip}. Such a well motivated interacting model has been investigated in \cite{Patil:2023rqy} where the dark matter field interacts with the dark energy quintessence. The model was analysed particularly in the presence of late time data sets along with CMB distance priors. In this work, we try to reanalyze this interacting model in the presence of full Planck likelihoods and the latest BAO observations from DESI. In addition, we also include observations from Type Ia Supernovae to obtain the cosmological constraints in the model.

The organization of this paper is as follows. In section \ref{model} we talk about the coupled quintessence model emphasizing on the underlying theory. In section \ref{meth_and_data}, we look at the cosmological data and the associated methods used in our analysis. We report the results and give the relevant discussions in section \ref{results}. Finally, we conclude our analysis and address the future possibilities in section \ref{summ}.

\section{Coupled Quintessence Model}
\label{model}
The Cosmological principle demands a homogeneous and isotropic universe on sufficiently large scales. The standard choice for the background metric that encapsulates this notion is given by the Friedmann–Lemaître–Robertson–Walker (FLRW) metric which can be expressed in spherical polar coordinates as,
\begin{equation}
    ds^2=g_{\mu\nu} dx^\mu dx^\nu =-dt^2+a(t)^2\left( \frac{dr^2}{1-kr^2}+r^2d\theta^2+r^2\sin^2\theta \ d\phi^2\right)
\end{equation}
Here $k$ is the spatial curvature constant which takes the values -1, 0, and 1 for open, flat, and closed universes respectively. $a(t)$ represents the scale factor of expansion and $g_{\mu\nu}$ denotes the metric tensor of the background spacetime. In our analysis, we will be dealing with a flat background spacetime, i.e. $k=0$. The action for our quintessence model consisting of the canonical scalar field $\phi$ can be written as, 
%\begin{comment}
%\begin{equation} \label{lang2}
%    S = \int d^4x \sqrt{-g} \left( \frac{R}{2\kappa} - \frac{1}{2} g^{\mu\nu} \partial_\mu \phi \partial_\nu \phi - V(\phi) + \mathcal{L}_{dm}(\chi, \phi) + \mathcal{L}_r +\mathcal{L}_b \right)
%\end{equation}
%\end{comment}
\begin{equation} \label{lang2}
    S = \int d^4x \sqrt{-g} \left( \frac{R}{2\kappa} - \frac{1}{2} g^{\mu\nu} \partial_\mu \phi \partial_\nu \phi - V(\phi) + \sum_i \mathcal{L}_m^i(\chi_i, \phi)
 \right).
\end{equation}
 Here $\kappa=1/M_{\rm Pl}^2$, with the reduced Planck mass $M_{\rm Pl}$ defined by, $M_{\rm Pl} \equiv \sqrt{\frac{1}{8\pi G}}$. Hereafter, the reduced Planck mass is considered as 1. $R$ denotes the Ricci scalar and $g$ represents the determinant of the metric tensor $g_{\mu \nu}$. $\mathcal{L}_m^i$ represents the lagrangian densities  for the different component fields that can include $\phi$ dependencies through the coupling.
 In this work, we are considering a non-gravitational coupling between the dark energy field and dark matter. The coupling of dark sector with baryons can be neglected since this is strongly constrained by the local gravity measurements \cite{Damour:1990eh}. The coupling with radiation can be eliminated due to conformal invariance. In \cite{Simpson:2010vh}, it is suggested that the dark components of the universe can have large scattering cross sections, with very few observational signatures. This provides a lot of freedom in considering the form of interaction in this class of models \cite{Wang:2016lxa}. The variation of the action given in eq. (\ref{lang2}) with respect to the metric tensor yields the stress-energy tensor, which can be decomposed as contributions coming from the dark energy and the matter parts  
\begin{equation}
    T_{\mu\nu} \equiv T_{\mu\nu}^{(\phi)} + T_{\mu\nu}^{(m)}
\end{equation}
Here $T_{\mu\nu}$ is the total energy momentum tensor, where as the quantities on the right hand side represent the respective components. Since both DE and DM are coupled they do not evolve independently. Nevertheless, they satisfy the local conservation equation together as 
\begin{equation}\label{conserv}
    -\nabla^{\mu}T_{\mu\nu}^{(\phi)} = Q_{\nu} = \nabla^{\mu}T_{\mu\nu}^{({\rm dm})}
\end{equation}
where $Q_{\nu}$ indicates the interaction between DE and DM components. The interaction term can be explicitly written as, 
\begin{equation}\label{inte}
    Q_{\nu} = F_{,\phi}\rho_{\rm dm}\nabla_{\nu}\phi
\end{equation}
where, $F_{,\phi} \equiv \frac{\partial F}{\partial \phi}$. The coupling strength $F(\phi)$ is given as $F_0e^{m\phi}$, with $m$ being a constant called the interaction parameter. DM is taken as a phenomenological fluid with energy density $\rho_{\rm dm}$ and pressure $P_{\rm dm}$. The case of cold dark matter is analyzed by taking $P_{\rm dm}$, or equivalently $\omega_{\rm dm}\equiv \frac{P_{\rm dm}}{\rho_{\rm dm}}$ to be 0. The uncoupled radiation and baryon components satisfy the local conservation equation independently as $\nabla^{\mu}T_{\mu\nu}^{(r)} = 0 = \nabla^{\mu}T_{\mu\nu}^{(b)} $. Now, we write the Friedmann equations as 
\begin{align}\label{fried}
3H^2 &= \frac{\dot{\phi}^2}{2} + V(\phi) + F(\phi)\rho_{\text{dm}} + \rho_{r} + \rho_b,\notag \\
3H^2 + 2\dot{H} &= -\left(\frac{\dot{\phi}^2}{2} - V(\phi) + F(\phi) P_{\text{dm}} + P_{r} + P_b\right).
\end{align}
where $\rho_{r,b}$ and $P_{r,b}$ denote the corresponding energy density and pressure for the radiation and baryon components, respectively. These energy densities are respectively scaled as $\rho_{r}^0a^{-4}$ and $\rho_{b}^0a^{-3}$ with the superscript 0 over the quantity representing its present day value.
From the conservation equation in eq. (\ref{conserv}), we get the continuity equations for the dark sector components as 
\begin{align}\label{conti}
\dot{\rho}_{\phi} + 3H \rho_{\phi}(1 + \omega_{\phi}) &= Q, \notag 
\\
\dot{\rho}_{\text{dm}} + 3H \rho_{\text{dm}}(1+\omega_{\rm dm}) &= -Q,
\end{align}
In order to understand the effect of this coupling on galaxy clustering and evolution of matter perturbations, we can look at the linear perturbation equations for this coupled model. We re-scale the DM density term ``$F(\phi)\rho_{\rm dm}$" as $\rho_c$. With this, the perturbation variable is defined as $\delta_c \equiv\frac{\delta \rho_c}{\rho_c}$, where $\delta \rho_c$ is the deviation from the background DM density $\rho_c$. While studying the cosmological perturbations, infinitesimal coordinate transformations of the form:
\begin{equation}\label{eq4.8}
    x^\mu \rightarrow \tilde{x}^\mu = x^\mu + \xi^\mu(x)
\end{equation}
with $\xi^\mu$ being an arbitrary four vector, induce changes in the perturbation variables that may not correspond to any physical alteration of the space-time. As a result, not all perturbations will be physically meaningful, with some of them being pure gauge artifacts \cite{Ajith:2025rty}. We can eliminate the gauge degrees of freedom by choosing a gauge. In this work, we analyse the linear perturbation equations in the synchronous comoving gauge with the line element: 
\begin{equation}
    ds^2 = a^2(\tau) \left(-d\tau^2 + (\delta_{ij} + h_{ij}) dx^i dx^j\right)
\end{equation}
where $h_{ij}$ denotes the perturbation in the metric tensor and $\tau$ indicates the conformal time. In this gauge, the velocity perturbation vanishes \cite{Ma:1995ey}. In this scenario, the adiabatic sound speed is equal to $c_s$, where $c_s^2 = \frac{d P}{d\rho}$, which is further equal to $\omega_{\rm dm}$ as it is a constant \cite{Ma:1995ey}. With the considered form of interaction parameter $Q$, the generalized DM density perturbation equation  can be written as \cite{valiviita2008large},
\begin{equation}\label{pert}
    \delta_{c}' = -(1+\omega_{\rm dm})\frac{h'}{2} + m\phi'\delta_{c}
\end{equation}
where the $'$ over the quantity represents its derivative with respect to the conformal time $\tau$.
%\subsection{Dynamical system Formulation}
It is convenient to study this system in terms of dimensionless variables. Hence, we use the dimensionless variables defined as \cite{Patil:2023rqy} 
\begin{equation}\label{EN}
    x = \frac{\dot\phi}{\sqrt{6}H},\quad y = \frac{\sqrt{V(\phi)}}{\sqrt{3}H}, \quad \Omega_{\text{dm}} = \frac {\rho_{\text{c}}}{3 H^2}, \quad \Omega_{r} = \frac{\rho_{r}}{3 H^2}, \quad \Omega_{b} = \frac{\rho_{b}}{3 H^2}
\end{equation}
Here, the dot over $\phi$ denotes its derivative with respect to time. In this formulation, the DE and DM density parameters are obtained as 
\begin{equation}\label{densp}
    \Omega_\phi = x^2+y^2,\quad \Omega_{\rm dm}= 1-(x^2+y^2)-\Omega_r - \Omega_b
\end{equation}
Similarly the DE Equation of State(EoS) parameter is given as 
\begin{equation}
    \omega_\phi \equiv \frac{P_{\phi}}{\rho_{\phi}} = \frac{x^2-y^2}{x^2+y^2}
\end{equation}
To simplify our equations in the analysis, we define another quantity: 
\begin{equation}
    \gamma = 1+\omega_\phi = \frac{2x^2}{x^2+y^2}
\end{equation}
and employ it in place of $\omega_\phi$ in our dynamical equations. The total (effective) EoS parameter is now given as, 
\begin{equation}
    \omega_{\rm eff} = (\gamma-1)\Omega_\phi +\frac{\Omega_r}{3} + \omega_{\rm dm}\Omega_{\rm dm}
\end{equation}
With this we obtain our set of autonomous equations as,
\begin{align} \label{dyncdm}
    \gamma' &= \left(2 - \gamma\right) \sqrt{\frac{3 \gamma}{ \Omega_{\phi}}} \left( - \sqrt{3 \gamma \Omega_{\phi}} + \lambda \Omega_{\phi} + m (1 - \Omega_{\phi} - \Omega_{r}-\Omega_{b}) \right), \notag \\
    \Omega_{\phi}' &=3(1-\gamma)\Omega_\phi(1-\Omega_\phi)+m\sqrt{3\gamma\Omega_\phi}(1-\Omega_\phi-\Omega_r-\Omega_b)+\Omega_\phi\Omega_r+3\ \omega_{\rm dm}\Omega_\phi(1-\Omega_b-\Omega_r-\Omega_\phi), \notag \\
    \Omega_{r}' &= \Omega_r(\Omega_r-1)+3\Omega_r\Omega_\phi(\gamma-1)+3 \ \omega_{\rm dm}\Omega_r(1-\Omega_b-\Omega_r-\Omega_\phi), \notag \\
    \Omega_b' &= \Omega_b(3\Omega_\phi(\gamma-1)+\Omega_r)+3 \ \omega_{\rm dm}\Omega_b(1-\Omega_b-\Omega_r-\Omega_\phi). 
\end{align}
Here the $'$ over the quantity denotes its derivative with respect to $N \equiv \ln \ a$. To reduce the dimensionality of the system, we choose an exponential potential of the form $V(\phi) \propto e^{-\lambda \phi}$, with $\lambda$ being a constant. 
\section{Methodology and Data}\label{meth_and_data}
We implement the coupled quintessence model in a modified version of the Boltzmann solver code \texttt{CAMB} \cite{Lewis:1999bs,2012JCAP...04..027H}, and perform Markov-Chain-Monte Carlo (MCMC) simulations using the publicly available tool \texttt{COBAYA} \cite{Torrado:2020dgo,2019ascl.soft10019T}. Using fourth order Runge-Kutta method for numerical integration, we solve the dynamical system given in eq. (\ref{dyncdm}) starting from $\ln a  = -7$, and evolving it to the present time. The initial conditions for the parameters $\gamma$, $\Omega_\phi$ and $\Omega_r$ are chosen to be $0.0001$, $1.2\times 10^{-9}$, $0.15$ respectively, in accordance with the values for the same obtained from a $\Lambda$CDM description. Since the effect of quintessence is negligible at $\ln a = -7$, we consider the initial value of the quintessence field $\phi$  to be 0. Our generic model has 9 free cosmological parameters. We constrain the baryonic matter density $\Omega_b h^{2}$ with a gaussian prior around 0.022 consistent with the constraints from Big Bang Nucleosynthesis (BBN). The other cosmological parameters: the dark matter density $\Omega_{\rm c}h^2$,  the logarithmic amplitude of primordial curvature spectrum $ {\rm{ln}}(10^{10} A_s)$ evaluated at a suitable pivot scale, $k = 0.05 \  \text{Mpc}^{-1}$ along with its tilt $ n_s $, the reionization optical depth $\tau_{\rm reio} $, the present value of the Hubble parameter $ H_0 $, the model parameters $\lambda$ and $m$, and the EoS parameter of DM $\omega_{\rm dm}$, are taken with uniform priors. The corresponding prior intervals for each of these parameters are given in Table (\ref{prior}). 
\begin{table}\label{data}
    \centering
    \begin{tabular}{|c|c|}
        \hline
        \hspace{3em} \textbf{Parameters} \hspace{3em} & \hspace{2.5em} \textbf{Prior Interval} \hspace{2.5em} \\
        \hline
        $ {\rm{ln}}(10^{10} A_s)$ & [1.61, 3.91] \\
        $n_s$ & [0.8, 1.2]\\
        $\tau_{\rm reio} $ & [0.01, 0.8] \\        
        $H_0$ & [20, 120]\\
       % $\Omega_bh^2$ & [a, b]\\
        $\Omega_{c}h^2$ & [0.001, 0.99]\\
        $\lambda$ & [0, $1.2$]\\
        $m$ & [0.01, 0.15]\\
        $\omega_{\rm dm}$ & [-0.01, 0.12]\\
        \hline
    \end{tabular}
    \caption{Prior intervals for the cosmological parameters used in the analysis}
    \label{prior}
\end{table}
We use the standard three-neutrino description with a massive neutrino of mass $m_{\nu}= 0.06$ eV, with the other two being  massless. The scenario of CDM is presented as a separate case with $\omega_{\rm dm}$ going to 0. In such a case, the number of cosmological parameters reduces by 1. The convergence of chains is ensured by having the Gelman-Rubin criterion $|R-1| \leq 0.05$ or the effective sample size becoming greater than $10^5$. We utilize  \texttt{GetDist} \cite{Lewis:2019xzd} and \texttt{BOBYQA} \cite{2018arXiv180400154C,2018arXiv181211343C} to analyze the chains and find the maximum likelihood posterior $\chi^2_{\rm MAP}$, respectively. We employ the current observational data to constrain our interacting model and obtain the best-fit values and the $1-2\sigma$ confidence level intervals for the parameters in our model.
In our analysis, we use the following publicly available datasets  
\begin{enumerate}
        \item \textbf{CMB:-} We utilize the CMB likelihood built from four distinct components. First, the small-scale ($\ell > 30$) temperature and polarization power spectra, $C_{\ell}^{TT}$, $C_{\ell}^{TE}$, and $C_{\ell}^{EE}$, derived from the Planck \texttt{CamSpec} likelihood \cite{Planck:2018vyg, Efstathiou:2019mdh, Rosenberg:2022sdy}. Second, the large-scale ($2 \leq \ell \leq 30$) temperature spectrum, $C_{\ell}^{TT}$, obtained from the  Planck \texttt{Commander} likelihood \cite{Planck:2018vyg,Planck:2019nip}. Third, the large-scale ($2 \leq \ell \leq 30$) E-mode polarization spectrum, $C_{\ell}^{EE}$, is provided by the Planck \texttt{SimAll} likelihood~\cite{Planck:2018vyg,Planck:2019nip}. Finally, the CMB lensing likelihood is included, making use of the high-precision reconstruction available from the NPIPE PR4 Planck dataset~\cite{Carron:2022eyg}. We denote this likelihood by Pl.

        \item \textbf{DESI:-} We use the 13 DESI-BAO DR2 (\cite{DESI:2025zgx,DESI:2025zpo}) measurements across the redshift range $0.1 < z < 4.2$ obtained from observations of about 14 million galaxies and quasars which include bright galaxy sample (BGS), luminous red galaxies (LRG), emission line galaxies (ELG), quasars (QSO), and Lyman-$\alpha$ tracers. These measurements are given in terms of the volume averaged distance $D_{\rm V}(z) / r_{\rm d}$, angular diameter distance $D_{\rm M}(z) / r_{\rm d}$ and comoving Hubble distance $D_{\rm H}(z) / r_{\rm d}$, where $r_{\rm d}$ is the sound horizon at the drag era. 
        %We have also included local $H_0$ measurements using Cepheid host galaxies from SH0ES dataset \cite{Riess:2020fzl}. 
        
        \item \textbf{SNeIa:-} We use the PantheonPlus (PP) dataset \cite{Scolnic:2021amr}, which contains 1701 light curves for 1550 spectroscopically confirmed Type Ia supernovae (SNeIa) covering the redshift range $0.001 < z < 2.26$. Additionally, we incorporate the Union3 compilation comprising 2087 SNe \cite{Rubin:2023ovl}. We also use the DESY5 sample comprising 1635 photometrically classified SNe from the released part of the full 5 year data of the Dark Energy Survey collaboration (with redshifts in the range 0.1 $< \ z\ <$ 1.3), complemented by 194 low-redshift SNe from the CfA3 \cite{2009ApJ...700..331H}, CfA4 \cite{2012ApJS..200...12H}, CSP \cite{Krisciunas:2017yoe}, and Foundation \cite{Foley:2017zdq} samples (with redshifts in the range 0.025 $< \ z\ <$ 0.1), for a total of 1829 SNe \cite{DES:2024jxu}. 
        %and the DES-Y5(DESY5) sample containing 1635 DES SNe spanning $0.10<z<1.13$ \cite{DES:2024jxu}.

        %\item \textbf{DES:-} Dark Energy survey (DES) data release 1 \citep{DES:2017myr} provides cosmological results from the analysis of galaxy clustering and weak gravitational lensing. This includes measurements from shear-shear, galaxy-galaxy,and galaxy-shear two-point correlation functions, referred to as ``$3\times2$ pt”, measured from 26 million source galaxies in four redshifts bins and 650,000 luminous red lens galaxies in five redshifts bins, for the shear and galaxy correlation functions.
        
\end{enumerate}

\section{Results and Discussion}\label{results}
\begin{figure}[h!]
    \centering    \includegraphics[width=0.5\linewidth,height=0.4\linewidth]{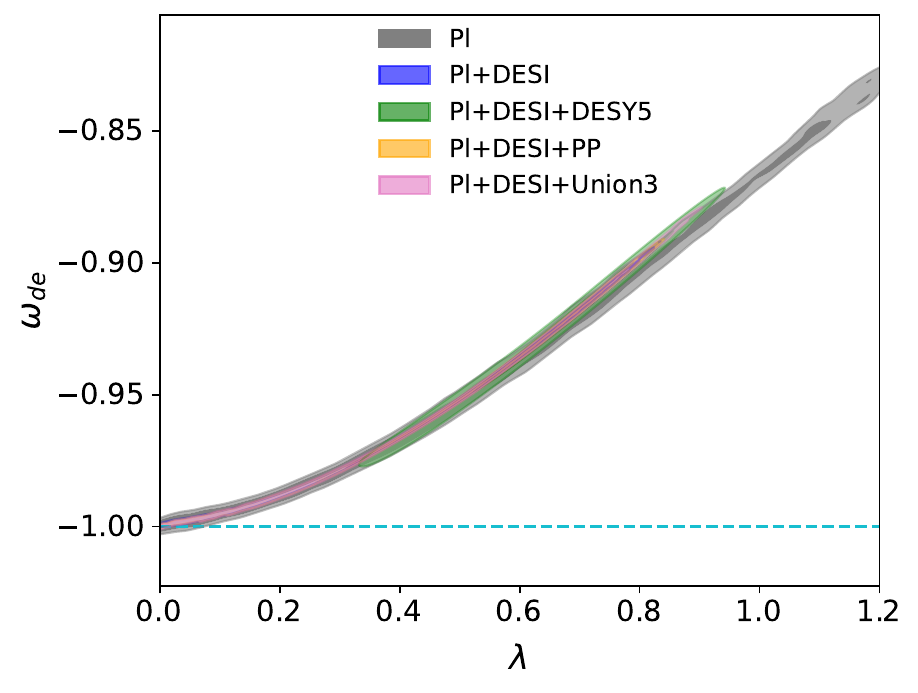}
    \caption{Variation of the Eos for dark energy, $\omega_{de}$, with the model parameter $\lambda$. The quintessence behavior of DE puts the lower bound on $\lambda$.}
    \label{fig:lambda}
\end{figure}

In this section, we analyze the constraints on the cosmological parameters in our model. We consider the $\Lambda$CDM, and coupled quintessence models in the presence of cold and warm dark matter.  The marginalized parameter constraints along with their $1-\sigma$ errors are reported in Table \ref{tab:model_comparison}. 
%Further, the $1\sigma$ and $2\sigma$ posterior distribution contours for the different cosmological parameters are depicted in figure \ref{fig:model_params}.
\begin{table}[ht!]
\centering
\resizebox{\textwidth}{!}{% Make the table fit the page width
\begin{tabular}{|l|l|c|c|c|c|c|c|c|c|}
\toprule
\hline
\textbf{Dataset} & \textbf{Model} & \textbf{$H_0$} & \textbf{$\Omega_m$} & \textbf{$\Omega_{\rm de}$} & \textbf{$\omega_{\rm dm}$} & \textbf{$\lambda$} & \textbf{$m$} & \textbf{$\omega_{\rm de}$} & \textbf{$\sigma_8$} \\
\midrule
\hline
% --- CMB Only Dataset ---
\multirow{3}{*}{{\footnotesize Pl}} 
& {\footnotesize $\Lambda$CDM} &  $67.24^{+0.45}_{-0.45}$  & $0.315^{+0.006}_{-0.006}$ & $0.684^{+0.006}_{-0.006}$ & -- & -- & -- & -- & $0.807^{+0.005}_{-0.005}$ \\
& {\footnotesize CQ+CDM} & $63.87^{+2.40}_{-1.19}$ & $0.349^{+0.013}_{-0.026}$ & $0.651^{+0.026}_{-0.013}$ & -- & $0.641^{+0.404}_{-0.419}$ & $0.064^{+0.013}_{-0.017}$ & $-0.924^{+0.079}_{-0.076}$ & $0.798^{+0.008}_{-0.007}$ \\
& {\footnotesize CQ+WDM} & $66.64^{+3.57}_{-4.43}$ & $0.323^{+0.036}_{-0.041}$ & $0.676^{+0.041}_{-0.036}$ & $0.0014^{+0.0014}_{-0.0023}$ & $0.515^{+0.174}_{-0.515}$ & $0.065^{+0.012}_{-0.012}$  & $-0.942^{+0.062}_{-0.058}$ & $0.811^{+0.017}_{-0.021}$ \\
\cmidrule(l){2-7} % Line under the model group
\hline
% --- CMB + DESI Dataset ---
\multirow{3}{*}{{\footnotesize Pl+DESI}} 
& {\footnotesize $\Lambda$CDM} &  $68.17^{+0.26}_{-0.29}$  & $0.302^{+0.003}_{-0.003}$ & $0.697^{+0.003}_{-0.003}$ & -- & -- & -- & -- & $0.805^{+0.005}_{-0.006}$\\
& {\footnotesize CQ+CDM} &  $67.34^{+0.27}_{-0.28}$ & $0.307^{+0.003}_{-0.003}$ & $0.693^{+0.003}_{-0.003}$ & -- & $0.099^{+0.025}_{-0.099}$ & $0.051^{+0.001}_{-0.003}$ & $-0.994^{+0.0005}_{-0.005}$ & $0.801^{+0.006}_{-0.007}$\\
& {\footnotesize CQ+WDM} &  $68.54^{+0.68}_{-0.54}$ & $0.302^{+0.004}_{-0.006}$ & $0.697^{+0.006}_{-0.004}$ & $0.0021^{+0.0004}_{-0.0005}$ & $0.356^{+0.119}_{-0.356}$ & $0.062^{+0.006}_{-0.010}$  & $-0.966^{+0.012}_{-0.033}$ & $0.820^{+0.006}_{-0.006}$ \\
\cmidrule(l){2-7}
\hline
% --- CMB + DESI + PantheonPlus Dataset ---
\multirow{3}{*}{{\footnotesize Pl+DESI+PP}} 
& {\footnotesize $\Lambda$CDM} &  $68.08^{+0.26}_{-0.27}$  & $0.303^{+0.003}_{-0.003}$ & $0.696^{+0.003}_{-0.003}$ & -- & -- & -- & -- & $0.805^{+0.005}_{-0.006}$\\
& {\footnotesize CQ+CDM} &  $67.27^{+0.29}_{-0.26}$ & $0.308^{+0.003}_{-0.003}$ & $0.692^{+0.003}_{-0.003}$ & -- & $0.110^{+0.029}_{-0.110}$ & $0.051^{+0.002}_{-0.003}$ & $-0.993^{+0.0009}_{-0.005}$ & $0.801^{+0.006}_{-0.007}$\\
& {\footnotesize CQ+WDM} &  $68.18^{+0.55}_{-0.56}$ & $0.305^{+0.005}_{-0.005}$ & $0.694^{+0.005}_{-0.005}$ & $0.0023^{+0.0005}_{-0.0005}$ & $0.491^{+0.237}_{-0.131}$ & $0.068^{+0.0095}_{-0.0068}$  & $-0.949^{+0.030}_{-0.026}$ & $0.819^{+0.007}_{-0.006}$ \\
\cmidrule(l){2-7}  
\hline
% --- CMB + DESI + Union3 Dataset ---
\multirow{3}{*}{{\footnotesize Pl+DESI+Union3}} 
& {\footnotesize $\Lambda$CDM} &  $68.09^{+0.27}_{-0.26}$  & $0.303^{+0.003}_{-0.003}$ & $0.696^{+0.003}_{-0.003}$ & -- & -- & -- & -- & $0.805^{+0.005}_{-0.006}$\\
& {\footnotesize CQ+CDM} &  $67.26^{+0.31}_{-0.26}$ & $0.308^{+0.003}_{-0.003}$ & $0.692^{+0.003}_{-0.003}$ & -- & $0.114^{+0.029}_{-0.114}$ & $0.050^{+0.002}_{-0.003}$ & $-0.993^{+0.0008}_{-0.009}$ & $0.801^{+0.005}_{-0.007}$\\
& {\footnotesize CQ+WDM} &  $67.98^{+0.62}_{-0.65}$ & $0.307^{+0.005}_{-0.005}$ & $0.693^{+0.005}_{-0.005}$ & $0.0023^{+0.0005}_{-0.0005}$ & $0.553^{+0.243}_{-0.123}$ & $0.070^{+0.010}_{-0.070}$  & $-0.939^{+0.035}_{-0.025}$ & $0.819^{+0.007}_{-0.006}$ \\
\cmidrule(l){2-7}
\hline
% --- CMB + DESI + DES Y5 Dataset ---
\multirow{3}{*}{{\footnotesize Pl+DESI+DES Y5}} 
& {\footnotesize $\Lambda$CDM} &  $67.99^{+0.26}_{-0.26}$  & $0.304^{+0.003}_{-0.003}$ & $0.695^{+0.003}_{-0.003}$ & -- & -- & -- & -- & $0.805^{+0.005}_{-0.006}$\\
& {\footnotesize CQ+CDM} &  $67.15^{+0.33}_{-0.26}$ & $0.309^{+0.003}_{-0.003}$ & $0.691^{+0.003}_{-0.003}$ & -- & $0.135^{+0.037}_{-0.135}$ & $0.051^{+0.002}_{-0.004}$ & $-0.992^{+0.002}_{-0.007}$ & $0.801^{+0.005}_{-0.006}$\\
& {\footnotesize CQ+WDM} &  $67.64^{+0.52}_{-0.50}$ & $0.310^{+0.004}_{-0.004}$ & $0.689^{+0.004}_{-0.004}$ & $0.0025^{+0.0004}_{-0.0005}$ & $0.667^{+0.144}_{-0.089}$ & $0.075^{+0.007}_{-0.005}$  & $-0.922^{+0.024}_{-0.018}$ & $0.819^{+0.007}_{-0.006}$\\
\hline
\bottomrule
\end{tabular}
}
\caption{The mean $\pm 1\sigma$ constraints on  cosmological parameters inferred from various datasets including DESI DR2, CMB, and supernovae and their combinations considered in this work. Here, $H_0$ is in units of km ${\rm s}^{-1}$ ${\rm Mpc}^{-1}$.}
\label{tab:model_comparison}
\end{table}
\begin{figure}[ht!]
    \centering
    % Left figure
    \begin{subfigure}{0.48\textwidth}
        \centering
        \includegraphics[width=1\linewidth]{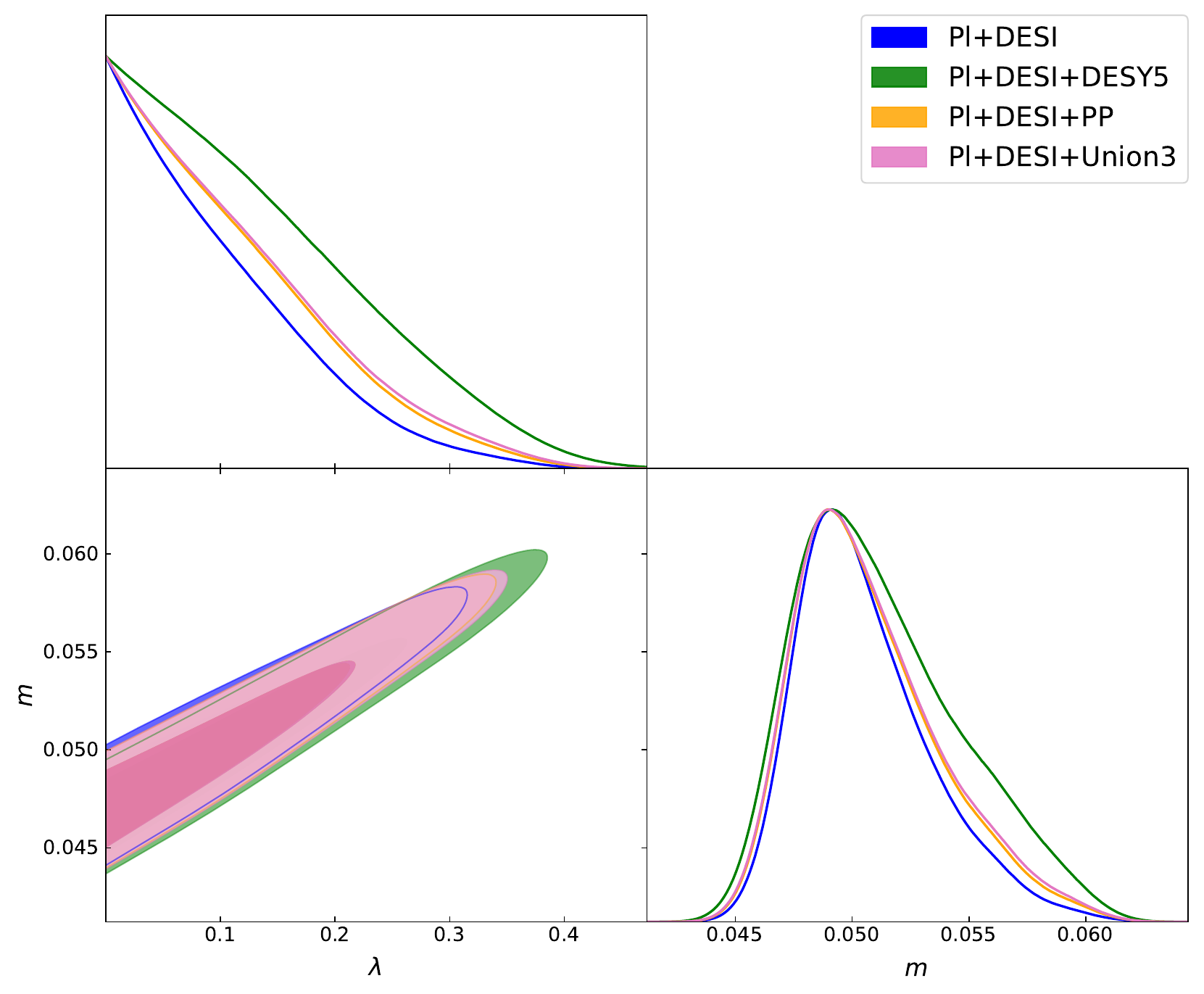}
        \caption{CQ+CDM model}
    \end{subfigure}
    \hfill
    % Right figure
    \begin{subfigure}{0.48\textwidth}
        \centering
        \includegraphics[width=\linewidth]{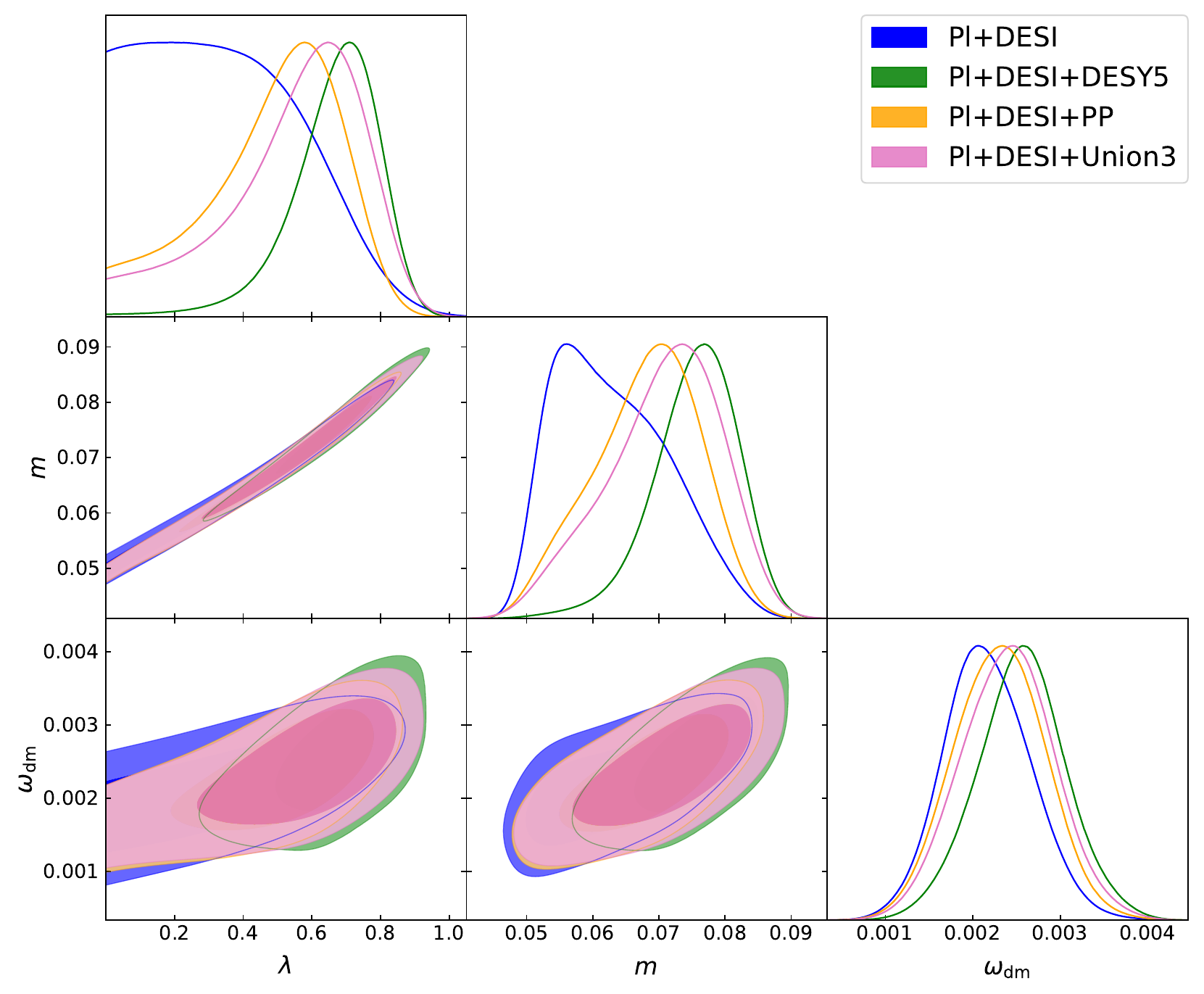}
        \caption{CQ+WDM model}
    \end{subfigure}
    
    % Common caption
    \caption{Corner plots of 1D and 2D marginalized posterior distributions for the coupled quintessence model in the presence of CDM and WDM, based on BAO from DESI DR2, CMB, and supernovae datasets. Contours at 68\% ($1\sigma$) and 95\% ($2\sigma$) levels
showing parameter constraints and correlations within the framework.}
    \label{fig:model_params}
\end{figure}
In fig. (\ref{fig:model_params}), we show the corner plot of the posterior distributions and parameter correlations derived from different
combinations of cosmological datasets, including CMB, DESI DR2, and various supernovae compilations (PP, Union3, and DESY5), for the coupled quintessence model in the context of both cold and warm dark matter. The diagonal panels display the
1D marginalized posterior distributions for each cosmological parameter, indicating the most probable values and associated uncertainties. The off-diagonal panels show the 2D joint posterior distributions between parameter pairs, with the inner and outer contours representing the 68\%
(1$\sigma$) and 95\% (2$\sigma$) confidence levels, respectively. The estimated values of $w_{\rm dm}$ are $0.0014^{+0.0014}_{-0.0023}$ (Pl), $0.0021^{+0.0004}_{-0.0005}$ (Pl+DESI), $0.0023^{+0.0005}_{-0.0005}$ (Pl+DESI+PP), $0.0023^{+0.0005}_{-0.0005}$ (Pl+DESI+Union3) and $0.0025^{+0.0004}_{-0.0005}$ (Pl+DESI+DES Y5) ), indicating the preference for a non-zero DM EoS parameter at $0.6\sigma$, $4.2\sigma$, $4.6\sigma$, $4.6\sigma$ and $5\sigma$
confidence levels, respectively. The model parameter $\lambda$ hits a wall assuming it can take only positive values. However, negative values of $\lambda$ can drive the equation of state parameter for the quintessence field  to go below -1, thereby crossing the phantom divide. Hence, in order to retain the quintessent behavior, the parameter $\lambda$ can only attain positive values. This behaviour is depicted in fig. (\ref{fig:lambda}).
We calculate the differences in $\chi^2_{\rm MAP}$ and deviance information criterion (DIC) to determine the preference of our model over the $\Lambda$CDM model. The DIC is calculated as 
\begin{equation}
    {\rm DIC} = 2\ \overline{\chi^{2}\left(\theta\right)} - \chi^{2}(\hat{\theta})
\end{equation}
with $\overline{\chi^{2}\left(\theta\right)}$ and $\chi^{2}(\hat{\theta})$ being the average of the effective $\chi^2$ over the posterior distribution and the best-fit $\chi^2$ respectively. The respective values for the dataset combinations we considered are given in table (\ref{tab:dic_chi2}). The difference in DIC, $\Delta{\rm DIC}$, is calculated by setting the standard $\Lambda$CDM model as the baseline, i. e. $\Delta{\rm DIC}_{\rm model}={\rm DIC}_{\rm model}-{\rm DIC}_{\Lambda{\rm CDM}}$. A negative $\Delta$DIC for a model indicates that the model is  preferred over the concordance $\Lambda$CDM model with the magnitude of $\Delta$DIC quantifying the degree of preference, while a positive $\Delta$DIC marks the opposite. For the CMB data, we can see that an interacting quintessence field in place of the cosmological constant can give negative values of $\Delta$DIC, indicating its preference over the $\Lambda$CDM model. But when other datasets are added along with the CMB dataset, this situation changes. With the inclusion of BAO data from DESI, the $\Delta$DIC is now positive for the coupled quintessence model, which indicates that it is less favored over the $\Lambda$CDM model. This trend continues even with the inclusion of supernovae data from PantheonPlus, Union3 and DES Y5 over the CMB+DESI combination. However, if we are to consider warm dark matter in place of the usual CDM, all the dataset combinations give out negative $\Delta$DIC values indicating the preference over the $\Lambda$CDM model. Particularly, when we are considering CMB+DESI+supernovae combinations, the $\Delta$DIC values are largely negative indicating a strong preference for the coupled quintessence model in the presence of warm dark matter.
\begin{table}[ht!]
    \centering
    \resizebox{9cm}{4.2cm}{
    \begin{tabular}{|l|l|c|c|}
    \toprule
     \hline
     \textbf{Dataset} & \textbf{Model}  & \textbf{$\boldsymbol{\Delta\chi^2_{\rm MAP}}$} &\textbf{$\boldsymbol{\Delta}$DIC}  \\
\midrule
\hline
      \multirow{3}{*}{Pl}  & $\Lambda$CDM  & 0 & 0 \\
         & CQ + CDM & -5.17 &  -3.34 \\
         & CQ + WDM \hspace{0.5em}  & -9.636  & -7.024  \\
         
         \hline
        \multirow{3}{*}{Pl + DESI}  & $\Lambda$CDM  & 0 & 0 \\
         & CQ + CDM & 6.468 &  11.366 \\
         & CQ + WDM  & -17.903  & -13.638  \\
         
         \hline  
         \multirow{3}{*}{Pl + DESI + PP}  & $\Lambda$CDM  & 0 & 0 \\
         & CQ + CDM & 5.353 &  10.092 \\
         & CQ + WDM  & \hspace{1em}-18.896\hspace{1em}  & \hspace{1em}-15.604\hspace{1em}  \\
         
         \hline   
         \multirow{3}{*}{Pl + DESI + Union3 \ }  & $\Lambda$CDM  & 0 & 0 \\
         & CQ + CDM & 5.456 &  9.849 \\
         & CQ + WDM  & -19.170  & -16.710  \\
         
         \hline
         \multirow{3}{*}{Pl + DESI + DES Y5 \ }  & $\Lambda$CDM  & 0 & 0 \\
         & CQ + CDM & 3.834 &  8.322 \\
         & CQ + WDM  & -24.299  & -20.120  \\
         
         \hline   
    \end{tabular}
    }
    \caption{Comparison of cosmological models using DIC with various dataset combinations. The baseline model for calculating the differences, ($\Delta\chi^2_{\rm MAP}$ and $\Delta$DIC), is taken to be the $\Lambda$CDM model for each dataset. A positive value of $\Delta$DIC indicates preference for the $\Lambda$CDM model.}
    \label{tab:dic_chi2}
\end{table}

However, the usage of Bayesian evidence as a selection criterion is superior to the usage of DIC, as the latter penalizes model complexity only weakly compared to Bayesian evidence. Hence, we also determine the Bayesian evidence for our model in the context of each dataset combination and compare it with the concordance model \cite{Li:2025dwz}. Here, we use the publicly available code \texttt{MCEvidence} \cite{Heavens:2017afc,Heavens:2017hkr} to compute the Bayes factor of the models. The Bayes factor is given by
\[
\ln B_{ij} = \ln Z_i - \ln Z_j
\]
in logarithmic space, where \(Z_i\) and \(Z_j\) are the Bayesian evidences of two models. Typically, we employ the Jeffreys scale \cite{Kass01061995,Trotta:2008qt} to gauge the strength of model preference: \(|\ln B_{ij}| < 1\) indicates inconclusive evidence; \(1 \leq |\ln B_{ij}| < 2.5\) indicates weak evidence; \(2.5 \leq |\ln B_{ij}| < 5\) indicates moderate evidence; \(5 \leq |\ln B_{ij}| < 10\) indicates strong evidence; and \(|\ln B_{ij}| \geq 10\) indicates decisive evidence. In Table (\ref{tab:evidence}), we display the Bayes factors $\ln B_{ij}$ for the coupled quintessence models, in comparison to the $\Lambda$CDM model, for the observational datasets we used in the analysis. Here, $i$ denotes the coupled quintesseence model, where as $j$ represents the baseline $\Lambda$CDM model. It is worth noting that negative values of the Bayes factor in this case indicate a preference for the $\Lambda$CDM model. Contrary to the results we obtained from the analysis on DIC, $\Lambda$CDM model is preferred over the coupled quintessence model even in the presence of warm dark matter. The $\Lambda$CDM model achieves a moderate preference over the quintessence scenario with CDM while considering CMB data from Planck. However, for all the other dataset combinations we analyzed, the quintessence field coupled with CDM is decisively disfavored. But when warm dark matter is considered, this disfavor decreases. In fact, for the CMB+DESI+DESY5 combination, the preference over $\Lambda$CDM can only be regarded weak, while the most of the other dataset combinations indicate a moderate evidence for favoring the $\Lambda$CDM model.   
\begin{table}[ht!]
    \centering
    \begin{tabular}{|l c c c c|}
    \hline
        \textbf{Datasets} & \hspace{5em}& \textbf{CQ+CDM} & \hspace{6em}& \textbf{CQ+WDM}\\
        %\hline
        \hline
         Pl&  & -2.66& & -5.86 \\
         Pl+DESI & & -12.54 & & -4.82 \\
         Pl+DESI+PP& & -11.75& & -4.17 \\
         Pl+DESI+Union3& & -11.71& & -3.82 \\
         Pl+DESI+DESY5& \hspace{3em} & -10.71& \hspace{3em}& -2.05 \\
         \hline
    \end{tabular}
    \caption{Summary of the $\ln B_{ij}$ values quantifying the evidence of the coupled quintessence model relative to the standard $\Lambda$CDM, utilizing the current observational datasets. A negative value indicates preference for $\Lambda$CDM.}
    \label{tab:evidence}
\end{table}
\section{Conclusion}
\label{summ}
In this work, we investigate a quintessence model for the dark energy where the dark sector consists of interactions between the dark energy field and the dark matter. We analyse this coupled quintessence model utilising the latest BAO
data from DESI DR2, CMB data from Planck, and supernovae data from  PantheonPlus, Union3 and DESY5. The observational constraints on the cosmological and model parameters are obtained in the context of different dataset combinations. To measure the goodness of fit of the model with regard to the standard $\Lambda$CDM model, we initially used DIC as a measure. We found that even though the coupled quintessence model has two extra parameters in comparison with the $\Lambda$CDM, the latter is preferred for all of the dataset combinations, except for the case of standalone CMB data. However, if we are to consider warm dark matter in place of its cold counterpart, this opens up a new parameter, making the coupled quintessence model strongly favored over the $\Lambda$CDM model across the various dataset combinations. On the contrary, if we are to compute Bayes factor for the models as a measure of model preference, $\Lambda$CDM takes preference even against the coupled quintessence model with warm dark matter. Further, Bayes factor estimation reveals that the coupled quintessence with CDM is decisively disfavored when CMB is used in combination with other datasets like BAO and supernovae. The inclusion of warm dark matter improves the goodness of fit of the model, but still fails to gain preference over the $\Lambda$CDM model. A possible direction of future analysis will be to expand this quintessence model to explore other forms of interaction that are linear ($Q\propto\rho_{\rm de}, \  Q\propto \rho_{\rm de}+\rho_{\rm dm}$) \cite{vanderWesthuizen:2025vcb}, nonlinear ($Q\propto H\frac{\rho_{\rm de}^2}{\rho_{\rm de}+\rho_{\rm dm}}, \ Q\propto H\frac{\rho_{\rm de} \ \rho_{\rm dm}}{\rho_{\rm de}+\rho_{\rm dm}}$) \cite{vanderWesthuizen:2025mnw} and those motivated from various Horndeski models \cite{Bansal:2024bbb}. Further, for the quintessence field, apart from the exponential potential, alternative forms such as a double exponential potential, and power-law potentials can be investigated \cite{Sen:2001xu,2015Ap&SS.356..383K}.  
\section*{Acknowledgements}
We thank Utkarsh Kumar and  Trupti Patil for the useful discussions. The numerical computations in this work was mostly carried out in the IISER Bhopal high performance computing clusters Raman and Bhaskara. We acknowledge the IISER Bhopal HPC facility for providing computing resources that have contributed to the research results reported in this paper.

\bibliographystyle{ieeetr}
\bibliography{refs}
\end{document}